\documentclass[twocolumn,showpacs,preprintnumbers,amsmath,amssymb,aps,nofootinbib]{revtex4}
\usepackage[cp1251]{inputenc}
\usepackage{color}
\usepackage{indentfirst}
\usepackage{amsmath,amsthm,amssymb}
\usepackage{graphicx}
\usepackage[active]{srcltx}
\usepackage{hyperref}
\usepackage{setspace}

\begin{document}

\title{ Impact of the strong  electromagnetic field on  the QCD  effective potential for  homogeneous  Abelian gluon field configurations}

\author{ Bogdan V. Galilo,  Sergei N.  Nedelko\footnote{nedelko@theor.jinr.ru}}
\affiliation{ Bogoliubov Laboratory of Theoretical Physics, JINR,
141980 Dubna, Russia, 
\\ and Department of Theoretical Physcis, Dubna International University, 141980 Dubna, Russia }

\begin{abstract}
The one-loop quark contribution to the QCD effective potential for the homogeneous Abelian gluon field in the presence of external strong electromagnetic field is evaluated. The structure of extrema of the potential as a function of the angles between chromoelectric, chromomagnetic  and electromagnetic fields is analyzed. 
In this setup, the electromagnetic field is considered as an external one while the gluon field represents domain  structured nonperturbative gluon configurations   related to the QCD vacuum   in the confinement phase. Two particularly interesting gluon configurations, (anti-)self-dual and crossed orthogonal chromomagnetic and chromoelectric fields,  are discussed specifically. Within this simplified framework it is shown  that the strong electromagnetic fields can play a catalysing  role for a deconfinement transition. At the qualitative level, the  present consideration can be seen as a highly  simplified study of an impact of  the electromagnetic fields generated in relativistic heavy ion collisions on the  strongly interacting hadronic matter. 

\end{abstract}


\maketitle

\section{Introduction}

The purpose of this paper is to study a potential  influence of the strong electromagnetic fields,  $eH\simeq\Lambda^2_{\mathrm QCD}$, on the QCD vacuum structure.  Electromagnetic fields with the strength of this order can emerge in relativistic heavy ion collisions. Before  proceeding, we have to decode our understanding of the  stock phrase "QCD vacuum structure``.   In pure gluodynamics a  physical vacuum can be characterized, first of all,  by two  invariants composed of  gauge field: scalar gluon condensate $\langle g^2F^2\rangle$ and pseudoscalar condensate $\langle g^2F\tilde F\rangle$ (for instance see discussion in \cite{Mink}).   Since parity is not broken in strong interactions,  the pseudoscalar condensate must be zero. Significance of the composite field $g^2F\tilde F$ becomes manifest in terms of topological susceptibility. In QCD with quarks another condensate $\langle m\bar \psi\psi\rangle$ comes into consideration. Identification of gauge field configurations which are  carriers of condensates and the method of their incorporation into the  formalism of quantum field theory can be seen as the most fundamental step towards  understanding  the mechanisms of confinement, chiral symmetry breaking and hadronization   in QCD.    This statement can be perceived  as  a kind of  platitude since implicitly  this step has to be assumed in all approaches dealing with configurations like center vortices, monopoles, instantons, etc.   However, the feeling of having just a commonplace here  relaxes  if one  identifies explicitly the point in the formalism where relevant condensates can be allowed or denied  to be nonzero.   As has been amphasized  in a recent paper~\cite{Faddeev0},  this  point can be  recognised in the choice of a functional space of  the gauge fields to be integrated over in the  QCD functional integral.  In the Euclidean functional integral approach to quantization of the pure YM theory  one starts with a symbol
\begin{eqnarray*}
 &&Z=N\int\limits_{{\cal F}} DA \exp\{-S[A]\},
 \end{eqnarray*}
 where the functional space ${\cal F}$  of fields is  subject to certain conditions, which can disable, in particular,  the gluon condensate (requirement of finite classical action $S[A]$, for instance)
 or enable it and  restrict the type of fields which can contribute to the condensates. The character of fields  in  ${\cal F}$ has to be defined self-consistently   on the basis of  quantum effective action. Enabling the gluon condensate means that gauge fields  $A_\mu^a$ should satisfy
 \begin{eqnarray}
 \label{cond0}
 {\cal F}=\{A: \lim_{V\to \infty}\frac{1}{V}\int_V d^4xg^2F^a_{\mu\nu} (x)F^a_{\mu\nu}(x) =B^2\}.
\end{eqnarray}
First of all,  the requirement of nonzero condensate $B^2\not=0$ singles  out  fields $B_\mu^a$ with the  strength which is constant almost everywhere in $R^4$, i.e. the part of $R^4$ where the field is inhomogenous  has measure (4-volume)  zero. The rest of deviations from homogeneity can be treated as  fluctuations in the background of $B_\mu^a$. Separation of the long range modes $B_\mu^a$ responsible for gluon condensate and the local fluctuations $Q_\mu^a$ in the background $B_\mu^a$, must be  supplemented by the
 gauge fixing condition. The background gauge condition $D(B)Q=0$ is the most natural choice. At the formal  level, the separation can be achieved by the insertion of identity
\begin{eqnarray*}
 1=\int\limits_{{\cal B}}DB \Phi[A,B]\int\limits_{{\cal Q}} DQ\int\limits_{\Omega}D\omega \delta[A^\omega-Q^\omega-B^\omega]
 \\
 \times\delta[D(B^\omega)Q^\omega],
\end{eqnarray*}
\begin{eqnarray}
\label{division}
 A_\mu^a=B_\mu^a+Q_\mu^a,
\end{eqnarray}
where $Q$ are   fluctuations of the gluon field with zero gluon condensate: $Q\in {\cal Q}$. Field $B_\mu^a$ are long range field configurations with, in general,  the nonzero condensate:  $B\in {\cal B}$. Performing the standard Faddeev-Popov procedure one arrives  at
\begin{eqnarray*}
Z &=&N'\int\limits_{{\cal B}}DB \int\limits_{{\cal Q}} DQ    \det[D(B)D(B+Q)]
 \\
 &&\times\delta[D(B)Q] \exp\{-S[B+Q]\}.
\end{eqnarray*}
The character of long-range fields has yet to be identified by the dynamics of fluctuations $Q$. At the formal level, integral over $Q$ defines an  effective action for the long range part of the gluon field
\begin{eqnarray*}
Z &=&N'\int\limits_{{\cal B}}DB\exp\{-S_{\rm eff}[B]\}.
\end{eqnarray*}
  Gluon fields $B_\mu^a$, which   correspond to the global minima of   $S_{\rm eff}[B]$,  dominate over the integral in the thermodynamic limit $V\to \infty$ and define the phase structure  of the  system. First of all, one has to take a look at fields with just constant  strength. There are two different kinds of this type of fields: Abelian covariantly constant fields $B_\mu^a=-\frac{1}{2}n^a B_{\mu\nu}x_\nu$ and non-Abelian constant vector potentials $B^a_\mu={\rm const}$. Unlike the former,   non-Abelian fields are unstable against small perturbations   $Q_\mu^a$ (for comprehensive discussion of the effective potential in pure Yang-Mills theory  see\cite{ Minkowski, Leutwyler}.    Pagels and Tomboulis studied an effective action for these fields within the context of scale anomaly \cite{Pagels} , Woloshyn and Trottier attempted lattice calculation \cite{Woloshin}. All these calculations indicated  a minimum of the effective action at nonzero Abelian (anti-)self-dual  field. Recently, the effective potential was calculated within the functional RG \cite{Pawlowski}. The result has also indicated a  minimum of the effective action at the nonzero Abelian (anti-)self-dual  field. In \cite{NG2011},   the Landau-Ginsburg Lagrangian for pure Yang-Mills gauge fields invariant under the standard space-time  and  local gauge  $SU(3)$  transformations was considered. It has been demonstrated that for $N_\texttt{c}=3$  a set of twelve  degenerated  minima of the action density exists as soon as a nonzero gluon condensate is postulated in the action. The minima are connected to each other by the Weyl group transformations associated with the color $su(3)$ algebra and  parity transformation. The presence of degenerated discrete minima in the Lagrangian leads to the solutions of the effective equations of motion in the form of the kink-like gauge field configurations interpolating between different minima.  The homogeneous field with a kink defect is the simplest  example of  gluon configurations which are homogeneous almost everywhere in $R^4$ and satisfy the basic condition Eq.(\ref{cond0}).  The spectrum of covariant derivative squared $D^2$  in the presence of the simplest solution, which interpolates between self-dual and anti-self-dual Abelian homogeneous fields,  was estimated.   This  kink configuration can be seen as a domain wall defect separating the regions with  self-dual and anti-self-dual Abelian gauge field. On the domain wall the gluon field is Abelian with orthogonal to each other chromomagnetic and chromoelectric fields.  For the aims of the present study it is important that the spectrum of   $D^2$ or $\not \!\!\!D$ in the (anti-)self-dual field  is purely discrete with bound state type eigenfunctions while for the crossed orthogonal fields the spectrum is continuous with the Landau level structure and the corresponding wave eigen functions.  
  
  The eigenvalues  and the square integrable eigenfunctions  of $D^2$ for the (anti-)self-dual field are
\begin{eqnarray*}
\lambda_{r} &= &4B\left( r + 1 \right)
\\
 \phi_{n m k l} (x) &=& C_{n m k l}\left(\beta_+^{+}\right)^{k} \left(\beta_-^{+}\right)^{l} 
 \left(\gamma_+^{+}\right)^{n}\left(\gamma_-^{+}\right)^{m} \phi_0(x),
 \\
  \phi_0(x)& =&  e^{-\frac{1}{2}Bx^2}, \  C_{n m k l}=\frac{1}{\sqrt{n!m!k!l!}\pi^2}, 
\end{eqnarray*}
where  $r=k+n $ for the self-dual \ field,  $r=l+n$  for  the anti-self-dual  field, $\beta_\pm^\pm$ and $\gamma_\pm^\pm$ are related to a set of creation and annihilation operators (details can be found in \cite{NG2011}). The spectrum is discrete.  In this background no color charged waves are enabled,
and there are no charged particle degrees of freedom. This is understood below as confinement of dynamical charged fields.

 Inside the  infinitely thin domain wall placed at  $x_1=0$ with the chromomagnetic field directed along the $y$ axis and the chromoelectric field along the $z$ axis   the
charged scalar field displays a continuous spectrum  similar to the Landau levels.  The eigen functions  square integrable over $x_3$  take the form
\begin{eqnarray*}
\label{Lpv}
 \phi_n(p_2,p_4|x_2,x_3,x_4)=\exp(-ip_4x_4-ip_2x_2)\chi_n(p_4|x_3),
\end{eqnarray*}
where the functions $\chi_n$ are
\begin{eqnarray*}
 \chi_n(p_4|x_3)=\exp\left\{-2\sqrt{2}B\left(x_3+\frac{p_4}{4B}\right)^2\right\}
 \\
 \times H_n\left(2^{3/4}\sqrt{B}\left(x_3+\frac{p_4}{4B}\right)\right).
\end{eqnarray*}
The eigenvalues look like
\begin{eqnarray*}
\lambda_n(p_2^2,p_4^2)=2\sqrt{2}B(2n+1)+p_2^2+p_4^2,
\end{eqnarray*}
 and correspond to the color charged quasiparticles with mass  $m_n^2=2\sqrt{2}B(2n+1)$  freely moving along the chromomagnetic field:
 \begin{eqnarray}
  p_0^2=p_2^2+m_n^2.
  \label{quasimn}
 \end{eqnarray}
  
  The purely discrete spectrum and bound state (four-dimentional oscillator) eigen functions can be treated as confinement of color charged fields in the (anti-)self-dual homogeneous field (in the bulk of $R^4$).   Landau levels and wave eigenfunctions   indicate the absence of confinement at the domain wall.   In other words,  charged particles are localized at the  wall. 
  
  It should be noted here that accurate separation of the specific Abelian part  $B$  of general vector potential $A$ as in Eq.(\ref{division}) 
 is a complicated problem.   The methods to tackle the problem were studied in \cite{Cho2,shabanov1,shabanov2,Faddeev,Kondo}.

For completeness,  we have to mention that in the context of center symmetry the dominance of the lumpy gauge field configurations was  discussed in  lattice calculations~\cite{Moran:2008xq, Moran:2007nc,deForcrand:2008aw,deForcrand:2006my,Ilgenfritz:2007ua}. In paper~\cite{deForcrand:2008aw}, an effective model of $SU(2)$ gauge theory for the domain wall formation was considered. 

 The idea of the  dominance of the gluon fields which are (anti-)self-dual Abelian almost everywhere turned out to be  phenomenologically efficient.
The model of confinement, chiral symmetry breaking and hadronization based on the ensemble of Abelian (anti-)self-dual fields was developed in a series of papers~\cite{EN,NK1,NK2}.  In the model, the direction of the gauge field in space and color space, and the duality of the field  are random parameters of the domains as well as pisitions of domain centers. All configurations of this type are summed up in the partition function.  The domain model exhibits confinement of static (square law) and dynamical quarks (absence of poles in the propagators of color charged fields, discrete spectrum of the corresponding differential operator), spontaneous breaking of the flavour chiral symmetry, $U_{\rm A}(1)$ symmetry is broken due to the axial anomaly,  strong $CP$ violation is absent in the model. With a minimal set of parameters (meson masses, gauge coupling constant, gluon condensate and mean domain size) the model gives rather accurate results for  meson masses from  all different parts of the spectrum: light mesons including excited states,  heavy-light mesons, heavy quarkonia).  The decay constants and some  form factors were also calculated within the model. The above mentioned  kink  configurations  have not been  yet   incorporated into the domain model directly  but strongly motivate  it.

\begin{figure}[h!]
 \centerline{\includegraphics[width=80mm,angle=0]{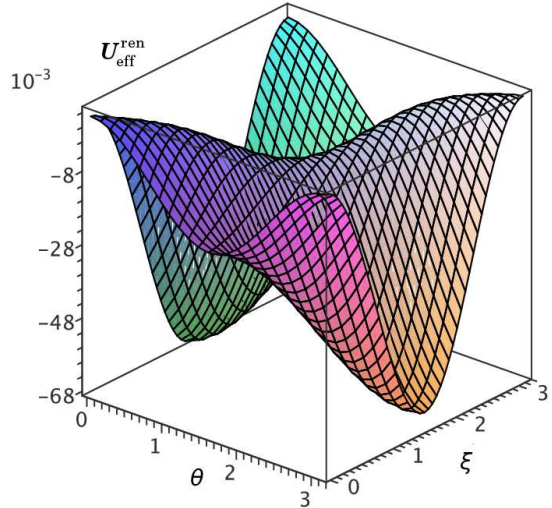}}
    \caption{Effective potential (in units of $B^2/8\pi^2$)  as a function of angles $\theta$ and $\xi$ for the pure magnetic field $H=.9B$ and  $\phi=\chi$ . The minimum is at $\theta=0$, $\xi=\pi/2$. }
 \label{Fig:H}
\end{figure}

Strong electromagnetic fields  can emerge in relativistic heavy ion collisions~\cite{Skokov:2009qp,toneev,Warringa}. Interplay of strong electromagnetic fields and nonperturbative gluon fields are expected to be important for understanding the dynamics of hadronic matter in heavy ion collisions. In particular,  these  fields can  initiate such phenomena as chiral and vortic magnetic effects  \cite{Warringa,Kharzeev2,Teryaev}.

In this paper, we study  an impact of the strong electromagnetic field on strong interactions in the context of lumpy or domain structured gluon fields.
The one-loop quark contribution to the QCD effective potential for the homogeneous Abelian gluon fields in the presence of   homogeneous electromagnetic field is evaluated. Extrema of the potential as a function of angles between chromoelectric, chromomagnetic  and crossed orthogonal electromagnetic fields are analysed. In this setup the electromagnetic field is considered as an external one  while the Abelian part of the gluon field represents domain  structured nonperturbative gluon configurations   related to  QCD    in the confinement phase.  It is shown that the quark contribution is minimal for the crossed  chromoelectric and chromomagnetic fields orthogonal to each other, which can be treated as a catalyzing
impact of strong electromagnetic fields on deconfinement in hadronic matter.  It should be stressed that this result has a very indirect relation to the real physics of heavy ion collision since it does not take into account the temperature and density effect. The present extremely simplified calculation can play  an instructive role for more realistic consideration. The main qualitative result of this paper is an observation that strong electromagnetic field could trigger a deconfinement transition in QCD.

\begin{widetext}

\section{One-loop quark contribution to the effective potential in the presence of arbitrary homogenous Abelian fields}

The one-loop contribution to the  QCD effective potential is defined by the Gaussian integral over quark fields $\psi$ with the covariant derivative which includes both the electromagnetic field and the Abelian homogeneous gluon field
\begin{eqnarray*}
 &&e^{-VU_{\rm eff}(G)}=\mathcal{N}\int D\psi D\bar\psi e^{ \int d^4x \bar\psi(x)\left( i \not  D-m \right)\psi(x)},
\\
&&U_{\rm eff}(G)= -\frac{1}{V}\ln\frac{\det( i\!\!\not\!\! D-m )}{\det( i\!\!\not\!\! \partial-m )}
\\
&&=\frac{1}{V}\int\limits_V d^4x \texttt{Tr}\int\limits^\infty_m d m' \left[ S(x,x|m')-S_0(x,x|m')\right],
\end{eqnarray*}
where $S(x,y|m)$ is the fermion propagator in external gauge fields with mass $m$. The following notation is used
\begin{eqnarray*}
&&\not\!\! D=\gamma_\mu D_\mu, \ \ \ \ D_\mu=\partial_\mu -iG_\mu, \ \ \ \ G_\mu=\hat B_\mu +qA_\mu,
\\
&&G_\mu =-\frac{1}{2}G_{\mu\nu}x_\nu,\ \ \ \ G_{\mu\nu}=qF_{\mu\nu}+ \hat n B_{\mu\nu},  
\\
&& G_{ij}=\varepsilon_{ijk}{\mathcal H}_{k},  \ 
 G_{4k}={\mathcal E}_k,
 \\
 &&\vec{\mathcal H}= \hat n \mathbf{H}_{\rm gl}+q\mathbf{H}, \ \ \ \  \vec {\mathcal E} =\hat n \mathbf{E}_{\rm gl}+q\mathbf{E}.
\end{eqnarray*}
Here electromagnetic fields are denoted as $\mathbf{H}$ and $\mathbf{E}$, and  $\mathbf{H}_{\rm gl}$ and $\mathbf{E}_{\rm gl}$ --  chromomagnetic and chromoelectric fields of the same value $\mathbf{H}_{\rm gl}=\mathbf{E_{\rm gl}}=B$, $q$ is quark electric charge. Trace includes  sum of the elements of diagonal matrices $\hat n$, $m$ and $q$. Two invariants of the gauge fields are
\begin{eqnarray*}
\mathcal{R} = \frac{1}{4}G_{\mu\nu}G_{\mu\nu} = \frac{1}{2}( \vec{\mathcal{H}}^2+ \vec{\mathcal{E}}^2), \ \ \ \
 \mathcal{Q} = G_{\mu\nu}\tilde G_{\mu\nu} = \vec{\mathcal{H}}\vec{\mathcal{E}}.
\end{eqnarray*}
The quark propagator can be calculated analytically (see appendix \ref{app:eff-potential}),
\begin{eqnarray*}
 S(x,y|m)=(m + i \!\not\!\! D_x)H(x,y|m),
\end{eqnarray*}
with
\begin{eqnarray}
 \label{H(x,y)}
&& H(x,y|m)
 =\frac{1}{m^2 + \!\not\!\! D^2}\delta(x-y)
 =e^{-\frac{i}{2}x_\mu G_{\mu\nu}y_\nu }   \frac{{\cal Q}}{16\pi^2}\int\limits_0^\infty  d s\frac{e^{-m^2s}}{\sinh(s\sqrt{{\cal Q}\sigma_-})\sinh(s\sqrt{{\cal Q}\sigma_+})}
\\\nonumber
&&
 \left[P_+\cosh(s| \vec {\mathcal{E}}- \vec {\mathcal{H}}|)+P_-\cosh(s| \vec {\mathcal{E}}+ \vec {\mathcal{H}}|)
  -\frac{1}{2}\sigma_{\mu\nu}[G_{\mu\nu}- \tilde G_{\mu\nu}]\frac{\sinh(s| \vec {\mathcal{E}}-\vec { \mathcal{H}}|)}{| \vec {\mathcal{E}}- \vec {\mathcal{H}|}}-\frac{1}{2}\sigma_{\mu\nu}[G_{\mu\nu}+\tilde G_{\mu\nu}]\frac{\sinh(s| \vec {\mathcal{E}}+\vec { \mathcal{H}}|)}{| \vec {\mathcal{E}}+\vec { \mathcal{H}}|}
 \right]
\\\nonumber
&&
\times\exp\left\{
 -\frac{\sqrt{{\cal Q}\sigma_+} \coth\left(s\sqrt{{\cal Q}\sigma_-}\right)  -  \sqrt{{\cal Q}\sigma_-} \coth\left(s\sqrt{{\cal Q}\sigma_+}\right)}  {4(\sigma_+ -\sigma_-)}
 (x-y)^2
\right.
\nonumber
\\
&&\left.
 -\frac{\sqrt{\cal Q\sigma_+}\coth\left( s\sqrt{{\cal Q}\sigma_+}\right)-\sqrt{\cal Q\sigma_-}\coth\left( s\sqrt{{\cal Q}\sigma_-}\right)}  {4\cal Q(\sigma_+ -\sigma_-)}
 G_{\mu}(x-y) G_{\mu}(x-y)
\right\},
\\
&&P_\pm =\frac{1}{2}(1\pm\gamma_5),
\ \ \sigma_{\mu\nu}=\frac{1}{2i}[\gamma_\mu,\gamma_\nu], 
\  \
 \sigma_\pm =\frac{\cal R}{\cal Q}\left(1\pm\sqrt{1-\frac{{\cal Q}^2}{{\cal R}^2}}\right).
 \nonumber
\end{eqnarray}
Using this propagator one gets for the effective potential
\begin{eqnarray}
\nonumber
&& U^{\rm ren}_{\rm eff}(G)=\texttt{Tr}\frac{\cal {Q}}{8\pi^2}
\int\limits_{s_0}^\infty  \frac{ds}{s}e^{-m^2s}
\frac{ \cosh(s| \vec {\mathcal{E}}- \vec {\mathcal{H}}|)+\cosh(s| \vec {\mathcal{E}}+ \vec {\mathcal{H}}|) }{\sinh(s\sqrt{{\cal Q}\sigma_-})\sinh(s\sqrt{{\cal Q}\sigma_+})}.
\label{ueff0}
\end{eqnarray}
Here $s_0$ regularises the UV divergence of the integral,   trace denotes summation over the elements  of the diagonal color matrix  $\hat n$  as well as quark charges  $q$  and  masses $m$ for all flavours under consideration.
Using the identities
\begin{eqnarray}
&& \cosh(s| \vec {\mathcal{E}}\pm \vec {\mathcal{H}}|)=\cosh(s(\rho_+\pm\rho_-)),
\ \ \sqrt{{\cal Q}\sigma_\pm}=\rho_\pm,
\label{idkappa}\\
&&\rho_\pm=\frac{1}{\sqrt{2}}\left(\sqrt{\mathcal{R}+\mathcal{Q}}\pm\sqrt{\mathcal{R}-\mathcal{Q}}\right),
\nonumber
\end{eqnarray}
one arrives at the renormalized effective potential written in terms of invariants $\mathcal{R}$  and   $\mathcal{Q}$  
\begin{eqnarray}
\nonumber
&&U_{\rm eff}(G)=U^{\rm ren}_{\rm eff}(G)+\delta U_{\rm eff},
\ \ \ 
\delta U_{\rm eff}= \frac{1}{8\pi^2} \left(\texttt{Tr}\frac{2}{3}\mathcal{R}\right)\int\limits_{s_0}^\infty \frac{ds}{s}e^{-\frac{m^2}{B}s},
\\
&& U^{\rm ren}_{\rm eff}(G)=\frac{B^2}{8\pi^2}
\int\limits_0^\infty  \frac{ds}{s^3}
\texttt{Tr}_n
\left[s\varkappa_+\coth(s\varkappa_+)
s\varkappa_-\coth(s\varkappa_-) -1 -\frac{s^2}{3}(\varkappa_+^2 +\varkappa_-^2)
\right]e^{-\frac{m^2}{B}s},
\label{uefff}
\end{eqnarray}
\begin{eqnarray}\nonumber
\varkappa_\pm
=\frac{1}{\sqrt{2}B}\left(\sqrt{\mathcal{R}+\mathcal{Q}}\pm\sqrt{\mathcal{R}-\mathcal{Q}}\right).
\end{eqnarray}
Through the identity (\ref{idkappa}) this expression can be reduced to the well known form of the effective potential, see \cite{Cho3} and references therein.

For the experimantal situation of heavy ion collisions, which we bear in mind,  it is sufficient to consider   the electric and magnetic fields orthogonal to each other and choose coordinate system with the $z$-axis along the magnetic field and the $x$-axis along the electric field,
\begin{eqnarray*}
 H_i=\delta_{i3}H,\ \ \ E_i=\delta_{i1}E, \ \ 
\mathbf{E}\mathbf{H} =0,
\end{eqnarray*}
For the physical electric field orthogonal to the magnetic field  invariants $\mathcal{R}$  and $\mathcal{Q}$  read
\begin{eqnarray}
\mathcal{R}&=&(H^2-E^2)/2+\hat n^2B^2+\hat nBH\cos(\theta)+iBE\cos(\chi)\sin(\xi),
\nonumber\\
\mathcal{Q}&=&\hat nBH\cos(\xi)+i\hat nBE\sin(\theta)\cos(\phi)+\hat n^2B^2(\sin(\theta)\sin(\xi)cos(\phi-\chi)+cos(\theta)cos(\xi)),
\label{RQ}
\end{eqnarray}
\end{widetext}
where $(\phi,\theta)$ are the spherical angles of the chromomagnetic field, and  $(\chi,\xi)$ are the spherical angles of the chromoelectric field in the chosen coordinate system.

\section{Minima of the effective potential}

For the pure magnetic field ($E=0$) the renormalised effective potential  is real. It depends on $\theta$, $\xi$ and the difference between $\phi$ and $\chi$.
Figure \ref{Fig:H} represents the effective potential for the case of three quark flavours, the masses are taken the same for all flavours. A minimum is achieved at 
$\theta=0$, $\xi=\pi/2$, i.e. for the crossed orthogonal chromomagnetic and chromoelectric fields.  The value of the potential does not depend on $\phi$ and $\chi$.  In other words, the chromomagnetic field is collinear to the magnetic field, and the chromoelectric field is orthogonal to the chromomagnetic field. The polar angle $\chi$ of $\mathbf{E}_{\rm gl}$ is not fixed.

\begin{figure}[h!]
 \centerline{\includegraphics[width=80mm,angle=0]{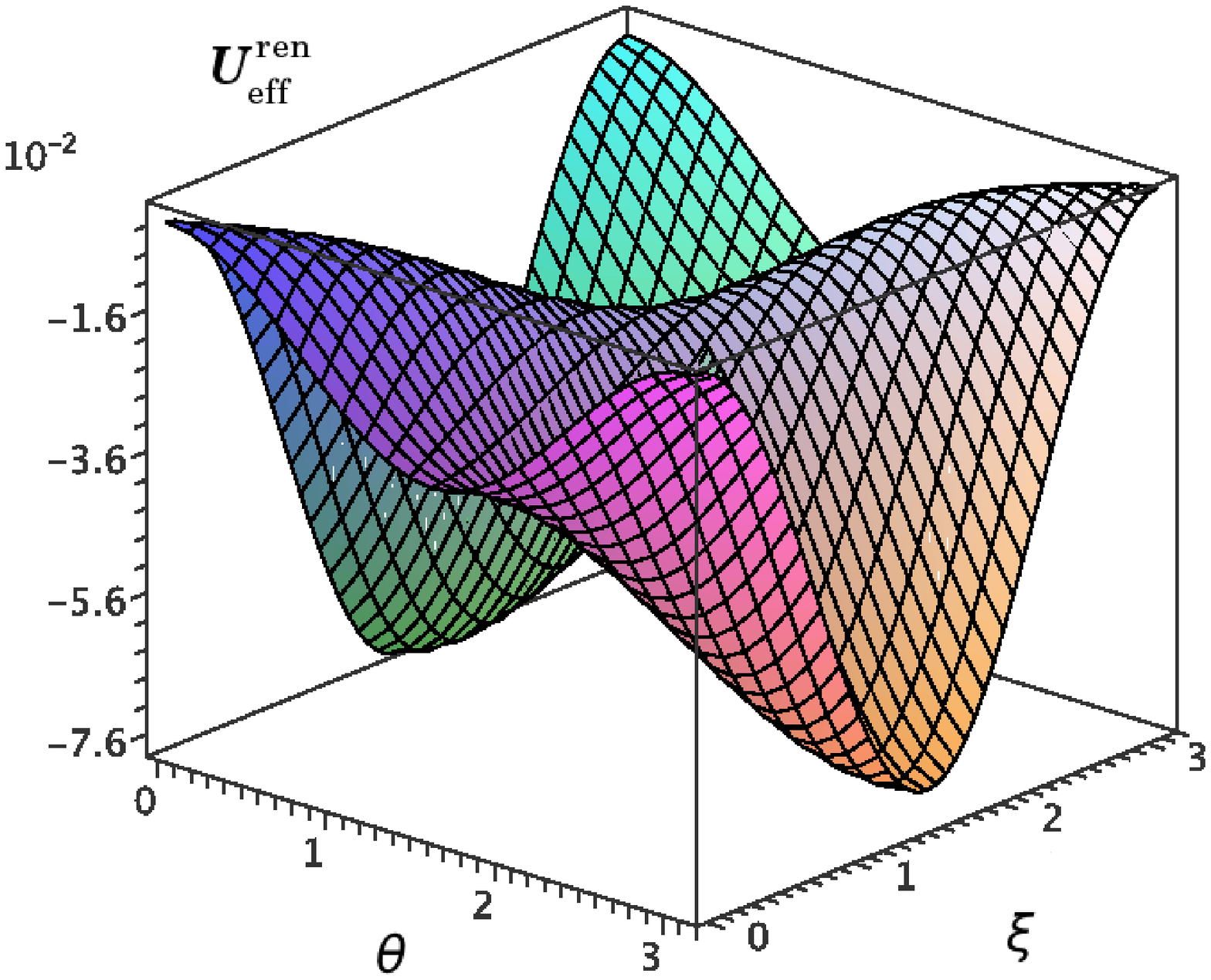}}
 \centerline{\includegraphics[width=80mm,angle=0]{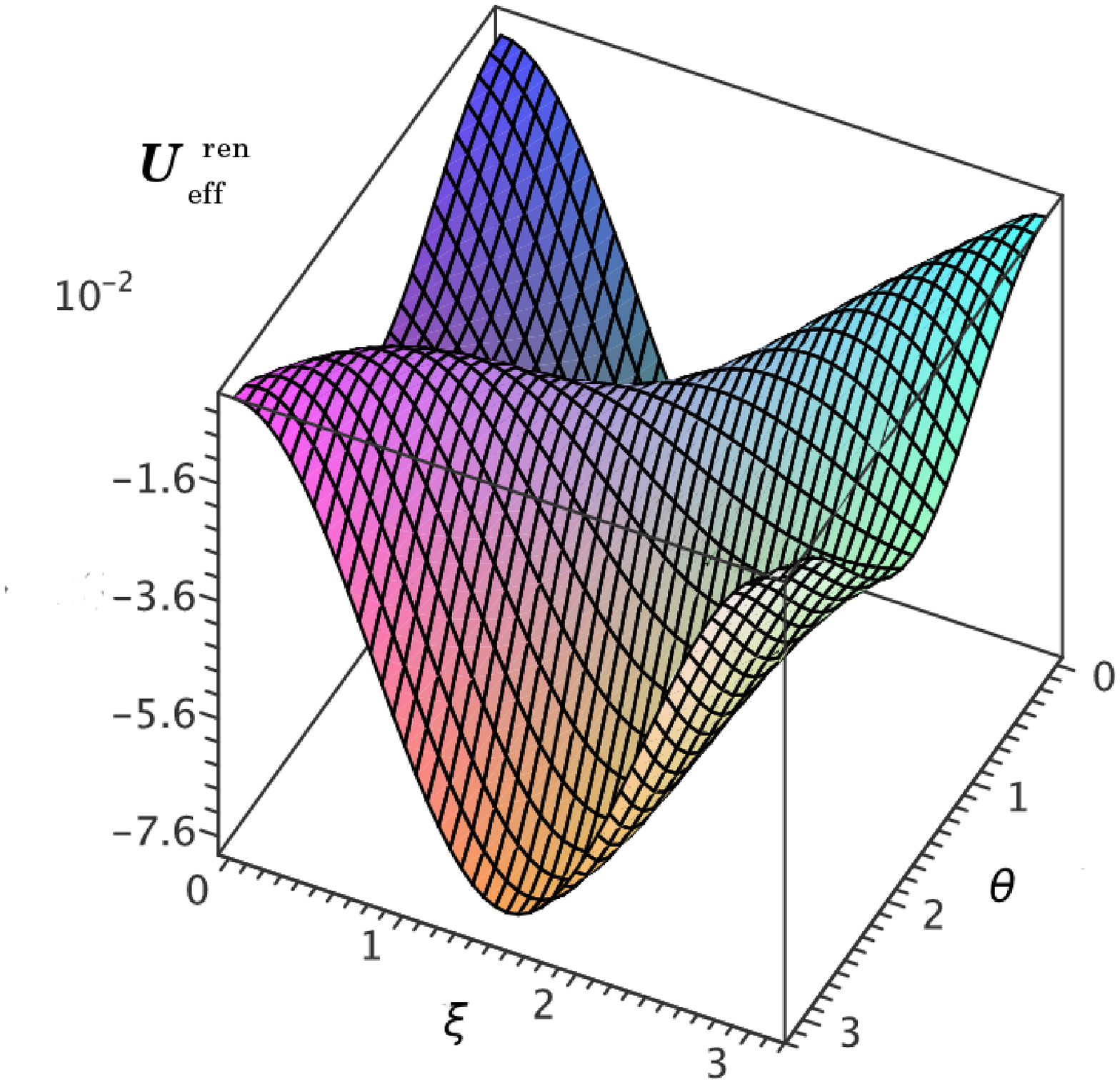}}
 \caption{Effective potential  (in units of $B^2/8\pi^2$)  for  the electric $E=.5B$ and  the magnetic $H=.9B$ fields as a function of the angles $\theta$ and $\xi$     ($\phi=\chi=\pi/2$ - upper figure, $\phi=\pi/2$, $\chi=3\pi/2$ - lower figure  ). }\label{Fig:EH}
\end{figure}

\begin{figure}[h!]
 \centerline{\includegraphics[width=80mm,angle=0]{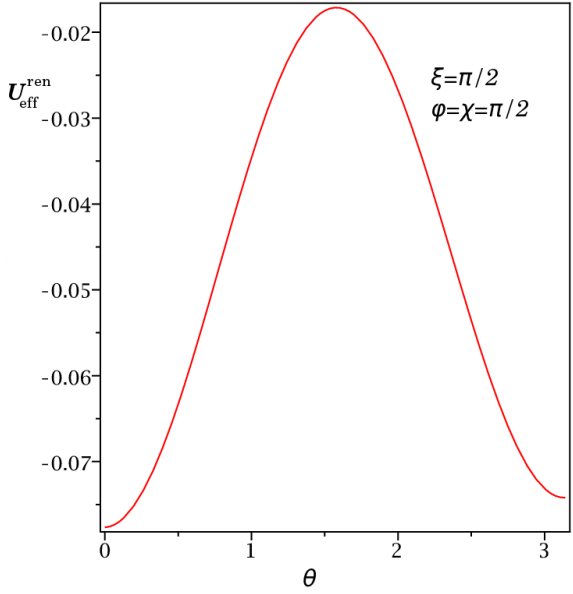}}
 \caption{Effective potential (in units of $B^2/8\pi^2$)  for  the electric $E=.5B$ and  the magnetic $H=.9B$ fields as a function of the angle $\theta$ and $\xi=\pi/2$     ($\phi=\chi=\pi/2$ ). }\label{Fig:EHxi}
\end{figure}

If the electric field is nonzero, the effective potential becomes comlex. The analytical properties of the effective potential as a function of complex $\cal{Q}$ and $\cal{R}$  can be studied by means of the convergent series representation of the integral (\ref{uefff}) obtained in \cite{Cho3}. The nonzero  imaginary part  of the potential would mean instability in the system.  However, the imaginary part does  vanish at the pont in the space of angles where the real part is minimal.  This  can be verified by inspection through the straightforward calculation of the real and imaginary parts of the potential.
The simplest way to find this stable minimum of the potential is to require that the RHS of Eqs.(\ref{RQ})  for $\mathcal{R}$ and $\mathcal{Q}$    to be real, which can be achieved by restriction imposed on the angles 
\begin{eqnarray*}
\sin(\xi)\cos(\chi)=0,
\\
\sin(\theta)\cos(\phi)=0.
\end{eqnarray*}
Figures \ref{Fig:EH} and \ref{Fig:EHxi}  illustrate that a globall minimum is achieved at $\theta=0$, $\xi=\pi/2$, $\chi=\pi/2$ or $3\pi/2$. It does not depend on $\phi$.  This minimum corresponds to the crossed orthogonal to each other chromoelectric and chromomagnetic fields, the chromomagnetic field  is collinear to the magnetic field, and the chromoelectric field is orthogonal to both magnetic and electric fields. The minimum exists for any value of the electromagnetic fields.  A  value of the potential at the minimum depends on their strengh.

\section{Discussion.}

We have analysed dependence of the one-loop quark contribution to the QCD effective action   on the
angles between  physical  homogeneous electromagnetic field and Abelian homogeneous Euclidean  gluon field.
The main result consists in observation that in the presence of the external  orthogonal magnetic and electric fields there exists a global mimium corresponding to the crossed orthogonal chromomagnetic and chromoelectric fields.  Unlike the (anti-)self-dual homogeneous gluon field minimising the  effective potential  in pure gluodynamics~\cite{Mink, Leutwyler, Pagels, Pawlowski}, the crossed orthogonal gluon field does not support  confinement of quarks:  the color charged quasi-particles do exist and can move along the direction of the magnetic field, see Eq.(\ref{quasimn}) . In this  sense, strong electromagnetic field  can  rearrange the structure of the global minima of the effective action of QCD.

If to take a liberty   to extrapolate  this  extremely simplified setup to the situation  of relativistic heavy ion collision, then one can expect that  a strong electromagnetic field generated during the collision  triggers  quark deconfinement  transition in hadronic matter.  Within the context of the domain model and the kink solution  one can think  that  a strong electromagnetic field produces a domain wall defect in the confining   gluon background exactly in the region where collision occurs. If so then deconfined quarks will move preferably along the direction of magnetic field but this will happen due to the gluon field configuration, due to QCD interaction, even after the switching the electromagnetic field off. However, prior to discussing the phenomenologically relevant observables arising from this effect, we should make the computation more realistic. First of all, this would imply incorporation of the temperature and baryonic density effects, and taking an inhomogeneity of the gauge fields into account as far as possible.

\section*{ACKNOWLEDGEMENTS}
 We acknowledge fruitful discussions with
Vyacheslav Toneev and Dmitri Pak. 

\appendix

\section{Quark propagator in the presence of arbitrary homogenous Abelian fields}
\label{app:eff-potential}
 The quark propagator 
\begin{eqnarray*}
&& S(x,y|m)=\frac{1}{m - i \!\not\!\! D_x}\delta(x-y)=(m + i \!\not\!\! D_x)H(x,y|m),
\\
&& H(x,y|m) =\frac{1}{m^2 + \!\not\!\! D^2}\delta(x-y),
\end{eqnarray*}
can be represented in the form
\begin{eqnarray}
\label{H1}
H(x,y) =
\int\limits_{0}^{\infty}ds \ e^{-m^2s} \ e^{-\frac{1}{2}s (\sigma G)}\ e^{s D^2}\delta(x-y),
\end{eqnarray}
with 
\begin{eqnarray*}
\!\not\!\! D^2& =&-D^2+\frac{1}{2}G_{\mu\nu}\sigma_{\mu\nu},
\\
 (\sigma G) &=& \sigma_{\mu\nu}G_{\mu\nu},  \ \  \sigma_{\mu\nu}=\frac{1}{2i}[\gamma_\mu,\gamma_\nu].
\end{eqnarray*}
Using  relations 
\begin{eqnarray*}
&& \left( \frac{1}{2}(\sigma G) \right)^{2n}
=( \mathcal{E}- \mathcal{H})^{2n}P_+ + ( \mathcal{E}+ \mathcal{H})^{2n}P_-,
\\
&& (\sigma G) P_\pm=\frac{1}{2}\sigma_{\mu\nu}[G_{\mu\nu}\mp \tilde G_{\mu\nu}],
\\
&&P_\pm =\frac{1}{2}(1\pm\gamma_5), \ \tilde G_{\mu\nu} = \frac{1}{2}\varepsilon_{\mu\nu\alpha\beta}G_{\alpha\beta}
\end{eqnarray*}
one can get
\begin{eqnarray}
\label{H2}
e^{-\frac{1}{2}s(\sigma G)}
&=& P_+\cosh(s| \vec{\mathcal{E}}- \vec{\mathcal{H}}|)+P_-\cosh(s| \vec{\mathcal{E}}+ \vec{\mathcal{H}}|)
\nonumber\\
&-&\frac{1}{2}\sigma_{\mu\nu}[G_{\mu\nu}- \tilde G_{\mu\nu}]\frac{\sinh(s| \vec{\mathcal{E}}- \vec{\mathcal{H}}|)}{| \vec{\mathcal{E}}- \vec{\mathcal{H}}|}
\nonumber\\
&-&\frac{1}{2}\sigma_{\mu\nu}[G_{\mu\nu}+\tilde G_{\mu\nu}]\frac{\sinh(s| \vec{\mathcal{E}}+ \vec{\mathcal{H}}|)}{| \vec{\mathcal{E}}+ \vec{\mathcal{H}}|}.
\end{eqnarray}

The integrand in Eq. (\ref{H(x,y)}) can be computed by means of the path integral representation,
\begin{eqnarray*}
&& \exp\{s D^2\}\delta(x-y)
=\int \delta a_\mu P_\beta
\exp\left\{
-\int\limits_0^1 d\beta a^2(\beta)
\right.
\\&&
\left.+2\sqrt{s}\int_0^1d\beta a_\mu(\beta)D_\mu
\right\} \delta(x-y)
\\
&=& \frac{1}{16\pi^2s^2}\exp\left\{-\frac{(x-y)^2}{4s}
-\frac{i}{2}x_\mu G_{\mu\nu}y_\nu \right\}\prod\limits_{k=1}^\infty \sqrt{ \det O(k)}
\\
&&\times \exp\left\{ \sum\limits_{n=1}^\infty
-\frac{2 s}{\pi^2n^2} G_\mu(x-y)O_{\mu\nu}^{-1}(n)G_\nu(x-y)\right\},
\end{eqnarray*}
where
\begin{eqnarray*}
 O_{\nu\alpha}(n)=\left[\delta_{\nu\alpha}+\frac{s^2}{\pi^2 n^2}  G_{\mu\nu}  G_{\mu\alpha}\right],
\end{eqnarray*}
and the determinant is
\begin{eqnarray*}
 && \det O(k)
 =\det\left[I-\frac{s^2}{\pi^2 n^2} G_{\mu\rho}G_{\rho\nu} \right]
\\
 && = \left[ 1+\frac{( \vec{\mathcal{E}}   \vec{\mathcal{H}})^2 s^4}{\pi^4k^4} + \frac{(  \vec{\mathcal{E}}^2+  \vec{\mathcal{H}}^2) s^2}{ \pi^2 k^2} \right]^2.
\end{eqnarray*}
With the notation
\begin{eqnarray*}
 &&{\cal Q}=( \mathcal{E}  \mathcal{H}),
 \ \ \ {\cal R}=\frac{1}{2} ( \mathcal{E}^2+ \mathcal{H}^2),
\\
 &&\sigma_\pm =\frac{\cal R}{\cal Q}\left(1\pm\sqrt{1-\frac{{\cal Q}^2}{{\cal R}^2}}\right),
 \ \ \ \sigma_+\sigma_-=1,
\end{eqnarray*}
one arrives at
\begin{eqnarray*}
 \det O(k) &=&\left\{  \left[ 1+\frac{{\cal Q}s^2\sigma_-}{\pi^2k^2} \right]  \left[ 1+ \frac{ {\cal Q}s^2\sigma_+}{\pi^2k^2} \right]   \right\}^2,
 \\
 \left[ \prod_{k=0}^\infty \det O^{1/2}(k)\right]^{-1}
 &=&\frac{s^2{\cal Q}}{\sinh(s\sqrt{{\cal Q}\sigma_-})\sinh(s\sqrt{{\cal Q}\sigma_+})}.
\end{eqnarray*}
\begin{widetext}
The sum in the exponent with $O_{\mu\nu}^{-1}$ can be evaluated as 
\begin{eqnarray*}
 \sum\limits_{n=1}^\infty \frac{2 s}{\pi^2n^2} G_\mu(x-y)O_{\mu\nu}^{-1}(n)G_\nu(x-y)
 = G_\mu(x-y)G_\mu(x-y) \Sigma_1 +(x-y)^2 \Sigma_2,
\end{eqnarray*}
where the terms $\Sigma_1$ and $\Sigma_2$ are given by
\begin{eqnarray*}
 \Sigma_1
 &=& \sum\limits_{n=1}^\infty \frac{2s\pi^2 n^2} {\pi^4n^4+2s^2\pi^2n^2{\cal R}+{\cal Q}^2s^4}
 = \frac{2\sigma_-s}{(\sigma_- - \sigma_+)}
 \sum\limits_{n=1}^\infty \frac{1}{\pi^2n^2+s^2{\cal Q}\sigma_-}
 +\frac{2\sigma_+s}{(\sigma_+ - \sigma_-)}\sum\limits_{n=1}^\infty \frac{1}{\pi^2n^2+s^2{\cal Q}\sigma_+}
\\
 &=& \frac    {\sqrt{{\cal Q}\sigma_+}\coth(s\sqrt{{\cal Q}\sigma_+}) - \sqrt{{\cal Q}\sigma_-}\coth(s\sqrt{{\cal Q}\sigma_-})}   {{\cal Q}(\sigma_+ - \sigma_-)}
\\
 \Sigma_2
 &=& \frac{1}{2} \sum\limits_{n=1}^\infty
     \frac{s^3{\cal Q}^2} {\pi^4n^4+2s^2\pi^2n^2{\cal R}+{\cal Q}^2s^4} 
 =\frac{s{\cal Q}} {2(\sigma_+ - \sigma_-)}\sum\limits_{n=1}^\infty  \left[ \frac{1}{\pi^2n^2+\frac{s^2{\cal Q}}{\sigma_+}} -  \frac{1}{\pi^2n^2+\frac{s^2{\cal Q}}{\sigma_-}}    \right]
\\
 &=& \frac{\sqrt{{\cal Q}\sigma_+}\coth\left(s\sqrt{{\cal Q}\sigma_-}\right) - \sqrt{{\cal Q}\sigma_-}\coth\left( s\sqrt{{\cal Q}\sigma_+}\right)}{4(\sigma_+ - \sigma_-)}
  - \frac{1}{4s}
\end{eqnarray*}
as a result of this calculation one gets  representation
\begin{eqnarray}
\label{H3}
&&\exp\{s D^2\}\delta(x-y)
=\frac{1}{16\pi^2}
  \frac{{\cal Q}}{\sinh(s\sqrt{{\cal Q}\sigma_-})\sinh(s\sqrt{{\cal Q}\sigma_+})}
  \exp\left\{-\frac{i}{2}x_\mu G_{\mu\nu}y_\nu \right\}
\nonumber\\
&&\times\exp\left\{-\frac{\sqrt{{\cal Q}\sigma_+} \coth\left(s\sqrt{{\cal Q}\sigma_-}\right)  -  \sqrt{{\cal Q}\sigma_-} \coth\left(s\sqrt{{\cal Q}\sigma_+}\right)}  {4(\sigma_+ -\sigma_-)}(x-y)^2
\right.
\nonumber\\
&&
\left.
 -\frac{\sqrt{\cal Q\sigma_+}\coth\left( s\sqrt{{\cal Q}\sigma_+}\right)-\sqrt{\cal Q\sigma_-}\coth\left( s\sqrt{{\cal Q}\sigma_-}\right)}  {4\cal Q(\sigma_+ -\sigma_-)}
 G_{\mu\nu}G_{\mu\rho}(x-y)_\nu(x-y)_\rho
\right\}.
\end{eqnarray}
Substitution of Eqs. (\ref{H2}) and (\ref{H3}) to Eq. (\ref{H1})  leads to  the quark propagator    (\ref{H(x,y)}).

\end{widetext}

\addtocontents{toc}{\protect\contentsline{section}{Bibliography}{\thepage}}

\end{document}